\documentclass[aps,prl,twocolumn,groupedaddress,showpacs,floatfix,superscriptaddress,longbibliography]{revtex4-2}
\usepackage[plainpages=false,pdfpagelabels,colorlinks=true,citecolor=blue,linkcolor=red,urlcolor=blue,citecolor=blue,pdftitle={Title},pdfauthor={},pdfdisplaydoctitle=true,pdfduplex=DuplexFlipLongEdge]{hyperref}
\usepackage{siunitx}
\usepackage{xcolor}
\usepackage{graphicx,color}
\usepackage{amsthm}
\usepackage{amsfonts}
\usepackage{algorithmic}
\usepackage{enumerate}
\usepackage{latexsym}
\usepackage{amsmath}
\usepackage{amssymb}
\usepackage{booktabs}
\usepackage{siunitx}

\usepackage{dcolumn}
\usepackage{bm}

\usepackage[T1]{fontenc}
\usepackage{mathptmx}

\begin{document} 
\title{Strain-enhanced edge ferromagnetism and bipolar magnetic semiconducting behavior in Janus graphene nanoribbons}
\author{Ran Liu}
\affiliation{School of Physics and Astronomy, and Key Laboratory of Multiscale Spin Physics (Ministry of Education), Beijing Normal University, Beijing 100875, China\\}
\author{Hongxing Liu}
\affiliation{School of Physics and Astronomy, and Key Laboratory of Multiscale Spin Physics (Ministry of Education), Beijing Normal University, Beijing 100875, China\\}
\author{ Junfeng Ren}
\affiliation{School of Physics and Electronics, Shandong Normal University, Jinan 250358, China\\}
\author{Tianxing Ma}
\email{txma@bnu.edu.cn}
\affiliation{School of Physics and Astronomy, and Key Laboratory of Multiscale Spin Physics (Ministry of Education), Beijing Normal University, Beijing 100875, China\\}

\begin{abstract}
Using first-principles density functional theory and determinant quantum Monte Carlo methods, we show that Janus graphene nanoribbons with topological defect arrays ($m=2$) exhibit robust intrinsic ferromagnetism across widths $W=2\text{--}6$, with bandgaps exceeding 200 $meV$ and stable ferromagnetic ground states. Notably, uniaxial tensile strain significantly enhances their ferromagnetic properties: at 25\% strain, the Curie temperature increases to $222K$—a fivefold improvement over unstrained systems and the highest reported for graphene-based nanoribbons. Strain also induces a reversible transition to a bipolar magnetic semiconductor, with spin-flipped valence and conduction band edges beyond 10\% strain. {\color{black}This dual functionality—strain-enhanced ferromagnetism and strain-induced spin flip—stems from strain-modulated $p_{z}$ orbital hybridization and strong direct exchange interaction.} Among these, $W=5$ Janus graphene nanoribbons emerge as potential candidates for room-temperature spintronic devices and strain-programmable quantum transport systems.
\end{abstract}

\date{Version 1.0 -- \today}

\maketitle

\section{Introduction}

It is well known that two-dimensional (2D) graphene and its derived materials are considered as promising candidates for electron transport devices due to their ultra-high carrier mobility and unique band structure \cite{C8NR00126J,C9NR08069D,Novoselov2005,PhysRevLett.100.016602}. However, pristine 2D graphene lacks intrinsic magnetism, which greatly limits its application in spin transport devices. In recent years, it has been shown that one-dimensional zigzag graphene nanoribbons (ZGNRs) can be obtained by cutting 2D graphene in a specific direction \cite{Li2012,C5RA06665D,PhysRevB.98.214204,C4CP03837A}. One-dimensional ZGNRs exhibit a critical advantage over 2D graphene: their bandgap and magnetic properties can be precisely modulated via width tuning, edge topological defects, chemical functionalization, or periodic pore engineering, positioning them as ideal candidates for advanced quantum transport and spin devices\cite{Kan2008,Mishra2020,10.1063/1.4747547,Gomes2025,PhysRevB.110.085141,doi:10.1126/science.abq6948,Blackwell2021,Ruffieux2016}. In addition, in the field of two-dimensional materials, the symmetry between the upper and lower layers can be broken by elemental doping or constructive defects to fabricate top and bottom asymmetric Janus materials. The emergence of Janus materials provides an attractive pathway for the study of topological materials, energy valley materials, and superconducting materials \cite{Zhang2013,Lu2017,Zhang2017,10.1063/5.0095203}.

Inspired by studies on 2D Janus materials, a recent work combined the concept of zigzag graphene nanoribbons with Janus structures\cite{Song2025}. Based on Lieb's theorem and topological classification theory \cite{PhysRevLett.62.1201,PhysRevLett.119.076401,Jiang2021}, two graphene nanoribbons with asymmetric zigzag edge were successfully synthesized experimentally.  This material is also known as Janus graphene nanoribbons(JGNRs). Introducing defects to one side eliminates magnetism on that side while preserving it on the opposite side in ZGNRs. This study enabled the ZGNRs to create ferromagnetic transport channels under one-dimensional conditions, providing an interesting  platform for realizing the preparation of one-dimensional ferromagnetic spin transport devices \cite{Song2025}.

In this paper, we first investigate the effect of JGNRs with different widths of defective benzene rings number of 2 ($m=2$) on their magnetic properties and electronic structures through first-principles calculations. Our goal is to understand how the widths of Janus graphene nanoribbons, together with edge defects, affect their electronic structure and magnetic characteristics. Next, we demonstrate a significant enhancement of the material's ferromagnetism by applying uniaxial tensile strain to the intrinsic structure. The Curie temperature of the material might reach a rare 222 $K$ at the limiting tensile strain (25\%), which makes JGNRs promising one-dimensional ferromagnetic spin-quantum devices operable near room temperature. Furthermore, our calculations demonstrate that JGNRs ($m=2$, $W=5$) act as strain-tunable bipolar magnetic semiconductors. This strain-mediated control of spin polarization offers an alternative to traditional magnetic-field modulation, suggesting a potential approach for developing spintronic devices with different operating modes.

\section{Model and Methods}

The structural and electronic properties of one-dimensional Janus graphene nanoribbons were calculated by means of the Vienna ab initio simulation package (VASP) based on density functional theory (DFT) \cite{PhysRevB.54.11169,KRESSE199615}. The ion-electron interaction is described by projector augmented wave (PAW) pseudopotential \cite{PhysRevB.50.17953}. The Perdew-Burke-Ernzerhof (PBE) functionalbased on generalized gradient approximation (GGA) is used to describe the exchange-correlation interactions \cite{PhysRevLett.77.3865}. The energy cutoff of the plane wave is defined as 500 $eV$ in this study. Each structure analyzed in the article undergoes full structural optimization, with the energies and forces convergence accuracies established at $10^{-6} eV$ and 0.001 $eV$/Å, respectively. The Brillouin zone is sampled using a $1 \times 7 \times 1\ $ Monkhorst-Pack (MP) k-point grid \cite{PhysRevB.13.5188}. Spin polarization were considered in all electronic structure calculations. To mitigate spurious interactions between periodic images, vacuum layers of \SI{23}{\angstrom} and \SI{15}{\angstrom} were imposed along the $a$- and $c$-axes, respectively. The Curie temperatures of the materials were calculated by the software Multi-dimensional Curie Temperature Simulation (MTC) developed by Zhang et al. based on Monte Carlo algorithm \cite{ZHANG2021110638}.

In addition, we performed additional simulations using the determinant quantum Monte Carlo (DQMC) method. This method is a reliable tool when studying the nature of magnetic correlations in electronic correlated system, which is especially significant when it comes to the change of energy band structure with strain and the change of edge topology. {\color{black}In our DQMC calculations, the lattice size is $7 \times 6 \times 2\ $, featuring a pristine zigzag edge on one side and a topological defect array of benzene motifs on the opposing edge.}

The Hamiltonian for a strained Janus graphene nanoribbon can be expressed as
\begin{equation}
    \begin{aligned}
        H = & \sum_{\mathbf{i},\eta,\sigma} t_{\eta} a_{\mathbf{i}\sigma}^{\dagger} b_{\mathbf{i} + \eta, \sigma} + \text{H.c.} 
            & + U \sum_{\mathbf{i}} \left( n_{\mathbf{ai}\uparrow} n_{\mathbf{ai}\downarrow} + n_{\mathbf{bi}\uparrow} n_{\mathbf{bi}\downarrow} \right) \\
            & + \mu \sum_{\mathbf{i},\sigma} \left( n_{\mathbf{ai}\sigma} + n_{\mathbf{bi}\sigma} \right).
    \end{aligned}
    \label{eq:hamiltonian}
\end{equation}
Here $ a_{\mathbf{i}\sigma} $ ($ a_{\mathbf{i}\sigma}^{\dagger} $) annihilates (creates) electrons at site $ \mathbf{R}_{\mathbf{i}} $ with spin $ \sigma $ ($ \sigma = \uparrow,\downarrow $) on sublattice A, as well as $ b_{\mathbf{i}\sigma} $ ($ b_{\mathbf{i}\sigma}^{\dagger} $) acting on electrons of sublattice B, $ n_{\mathbf{a}\mathbf{i}\sigma} = a_{\mathbf{i}\sigma}^{\dagger}a_{\mathbf{i}\sigma} $ and $ n_{\mathbf{b}\mathbf{i}\sigma} = b_{\mathbf{i}\sigma}^{\dagger}b_{\mathbf{i}\sigma} $. $ U $ is the on-site Hubbard interaction and $ \mu $ is the chemical potential. On such honeycomb lattice, $ t_{\eta} $ denotes the nearest-neighbor hopping integral. {\color{black}The hopping parameter is configured as $t_\eta(\varepsilon)= t_0 e^{-3.37(l/d-1)}$
, where $l=\frac{a}{2}/3\sqrt{3+9{{\left( 1+\varepsilon  \right)}^{2}}}$, and ${{t}_{0}}$ = 1, where a is the distance between the centers of the two hexagonal lattices, d is the carbon-carbon bond length, and $\varepsilon$ is the magnitude of uniaxial tensile strain.}

In DQMC, the basic strategy is to represent the partition function as a high-dimensional integral over a set of auxiliary fields, and subsequently complete the integration operation by Monte Carlo techniques. The current simulation uses 8000 scans to bring the system to an equilibrium state and an additional 30000 scans, each of which generates a measurement. These measurements are divided into ten intervals used to construct a coarse-grained mean basis, and error estimates are based on the standard deviation from the mean. Specific technical details can be found in the literature \cite{PhysRevD.24.2278,PhysRevLett.108.066402,PhysRevB.91.075410,PhysRevB.110.235128,2025-0862,PhysRevB.110.085103,PhysRevLett.120.116601,Chen2024}.

\section{Results and Discussion}

\begin{figure}[ht]
    \centering
    \includegraphics[width=0.5\textwidth]{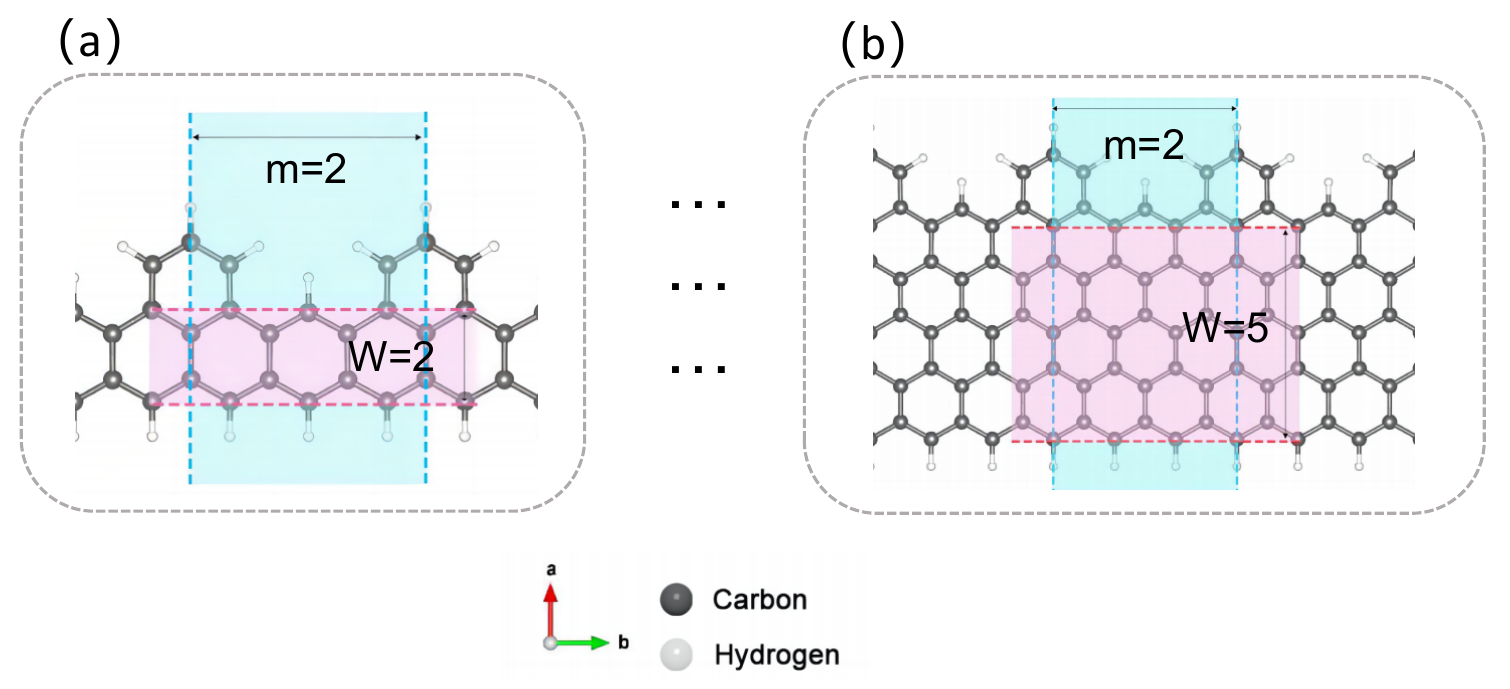} 
    \caption{(a) Shows Janus graphene nanoribbons (JGNRs) with $m = 2$ and $W = 2$. (b) Shows JGNRs with $m = 2$ and $W = 5$. Here, m represents the number of defective benzene rings and W represents the width of the nanoribbon.}
    \label{fig:1}
\end{figure}

\begin{figure}[ht]
    \centering
    \includegraphics[width=0.45\textwidth]{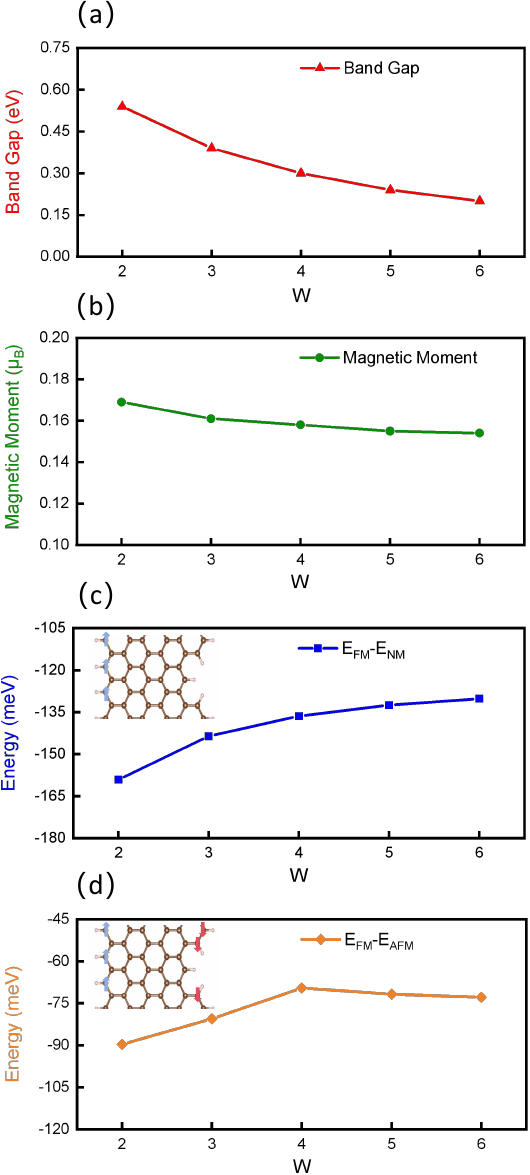} 
    \caption{(a) Total bandgap, (b) local magnetic moment at zigzag edge, (c) Energy difference (FM-NM), and (d) Energy difference (FM-AFM) as a function of JGNRs width. The inset figures in panels (c) and (d) are schematic diagrams of the FM and AFM states, respectively.}
     \label{fig:2}
\end{figure}

Fig.\ref{fig:1} illustrates the crystal structure of JGNRs optimized for a zigzag edge topology with the number of defective benzene rings of 2 $(m=2)$ and widths of 2, 3, 4, 5, and 6 $(W=2, 3, 4, 5, 6)$.  Song et al. firstly theoretically designed the structure based on Lieb's Theorem and Topological Classification Theory, and then synthesized the structures experimentally for the number of defective benzene rings of 2 ($m= 2$) and  widths of 4 and 5 respectively\cite{Song2025}. We constructed JGNRs structures with $m=2$ and widths of 2, 3, and 6 through theoretical modeling to investigate the impact of different widths on the electronic structure of this material.

According to the chiral phase index and the Jiang-Louie formula \cite{Jiang2021}, firstly, we can define the sign function $R = -((m-2)/(m+1))$, and when $m = 1$ in the formula is that $R > 0$, JGNRs exhibit identical spin alignment at two edges with different spin magnitudes, resulting in a ferromagnetic ground state. On the contrary, when $m > 2$ and $-1 < R < 0$, the two edges of the JGNRs show spin arrangements in opposite directions and different spin intensities, forming a subferromagnetic ground state. It is worth noting that when $m = 2$ and $R = 0$, no fringe state exists at the defect edge, resulting in the ferromagnetic ground state appearing only at the zigzag edge on the other side. For conventional ZGNRs, the extreme case of $m = \infty (R = -1)$ can be considered. At this point, the perturbation caused by infinitely spaced benzene rings vanishes, and the system exhibits an antiferromagnetic ground state. By this artificial introduction of a topological defect structure, the $Z_{2}$ symmetry of the spins on the boundaries of the ZGNRs is broken, giving rise to a ferromagnetic boundary state that exists only at the zigzag edge. Subsequent experimental results proved that the theoretical prediction is completely correct \cite{Song2025}.


In our calculations, it is observed that the lattice constants of the five different JGNRs are all determined to be 7.42 Å. This consistent lattice constant is attributed to the fact that variations in the widths of the JGNRs solely affect the number of atoms and the atomic arrangement along the non-periodic direction (a-direction), while the width remains constant at $m = 2$ along the periodic direction (b-direction). In order to eliminate dangling bonds caused by unsaturation of edge carbon atoms and maintain structural stability, we passivated all edge carbon atoms with hydrogen atoms.

After determining the crystal structure, we first calculated the magnetic ground states of the JGNRs with different widths, and it can be seen in Fig.\ref{fig:2}(c) that the energy of the ferromagnetic state is always lower than that of the nonmagnetic state at any width, and thus the ground states of the JGNRs are all ferromagnetic when $W$ = 2, 3, 4, 5, and 6.

Fig.\ref{fig:2}(a)--(c) demonstrate that JGNRs with $W = 2\text{--}6$ maintain significant spin polarization while exhibiting bandgaps consistently exceeding 200~\text{meV}. Notably, increasing JGNRs width reduces bandgap (Fig.\ref{fig:2}(a)) and enhances density of states in the interval $-2 \sim 2~\text{eV}$. This behavior arises from that the perturbative interactions between electrons of different atoms gradually increase with the increase of the number of atoms in the a direction. Such enhanced interactions drive non-magnetic splitting of high degenerate carbon energy levels, inducing band structure compaction. Simultaneously, the energy bands near the Fermi level with spin-down and spin-up states exhibit a gradually decreasing band gap as the width increases, due to the enhanced orbital interaction.

\begin{figure}[ht]
    \centering
    \includegraphics[width=0.45\textwidth]{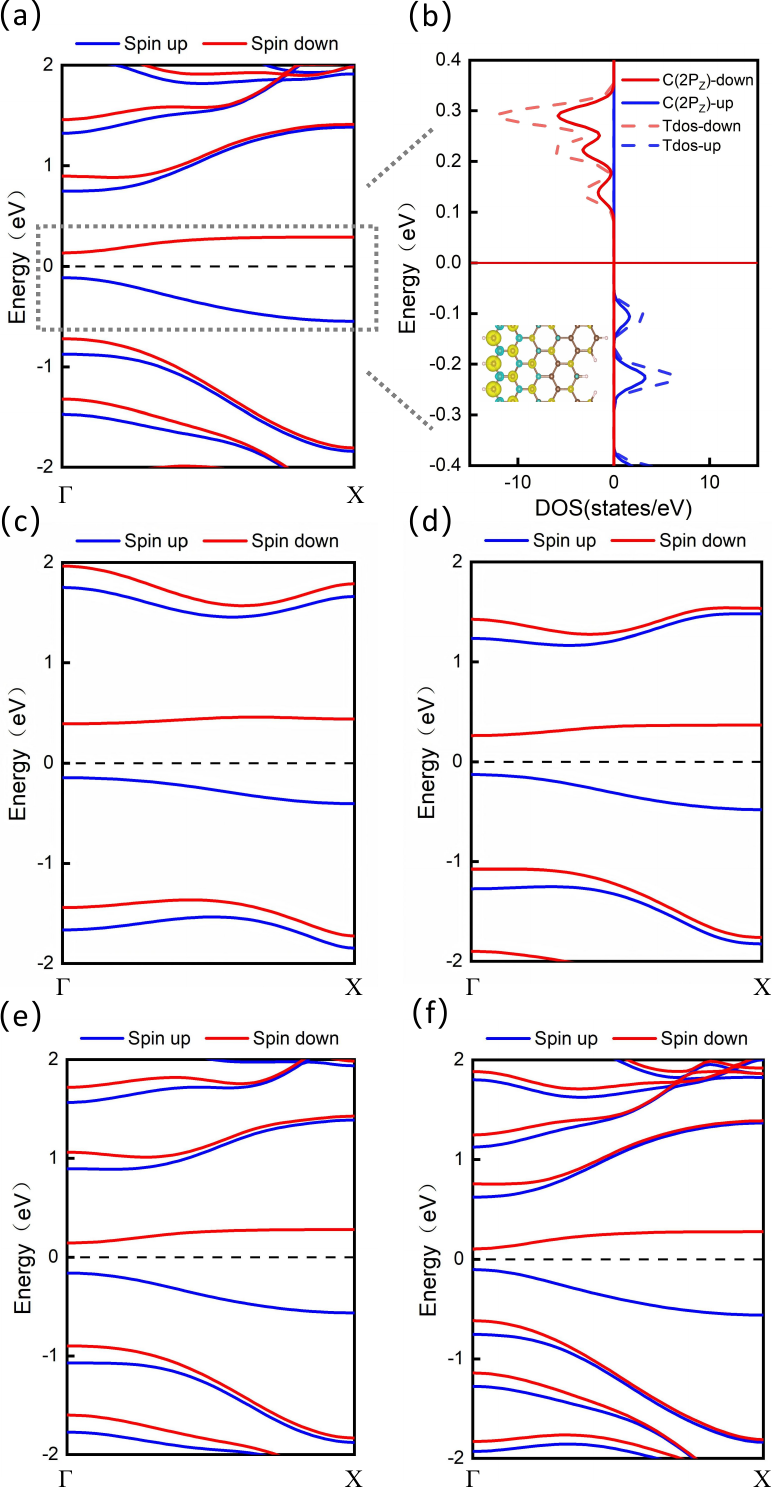} 
    \caption{(a) and (b) The spin-polarized energy bands and spin-polarized density of states of the JGNRs for $m = 2$ and $W = 5$, respectively, where the inset of (b) shows the spin charge density of the corresponding structure. (c)-(f) Spin-polarized energy bands of the JGNRs for $m = 2$, $W = 2,3,4,6$, respectively.}
    \label{fig:3}
\end{figure}

From Fig.\ref{fig:2}(b), we can see that the local magnetic moments of the carbon atoms at the zigzag edge boundaries decrease slightly with the increase JGNRs of width. Simultaneously, a slight reduction in magnetic moment attenuates the magnetic exchange coupling among carbon atoms, consequently diminishing the energy difference between ferromagnetic and nonmagnetic states, these interpretations align with the result in Fig.\ref{fig:2}(c). {\color{black}As shown in Fig.\ref{fig:2}(d), the energy difference between the ferromagnetic and antiferromagnetic states remains negative, demonstrating that JGNRs maintain a ferromagnetic ground state regardless of their width, without undergoing a transition to an antiferromagnetic state.} Therefore, through our study, we show that for $m = 2$ JGNRs, more materials with different widths can be synthesized experimentally for the study of their transport properties and magnetic properties without any effect on the intrinsic properties of their ferromagnetic ground states. {\color{black}Additionally, we calculated the magnetic ground state of the pristine carbon lattice after removing hydrogen dangling bonds and found it to remain ferromagnetic. This confirms that the ferromagnetism observed in JGNRs is an intrinsic property of the material rather than a result of hydrogen passivation.}

\begin{figure}[ht]
\centering
\includegraphics[width=0.45\textwidth]{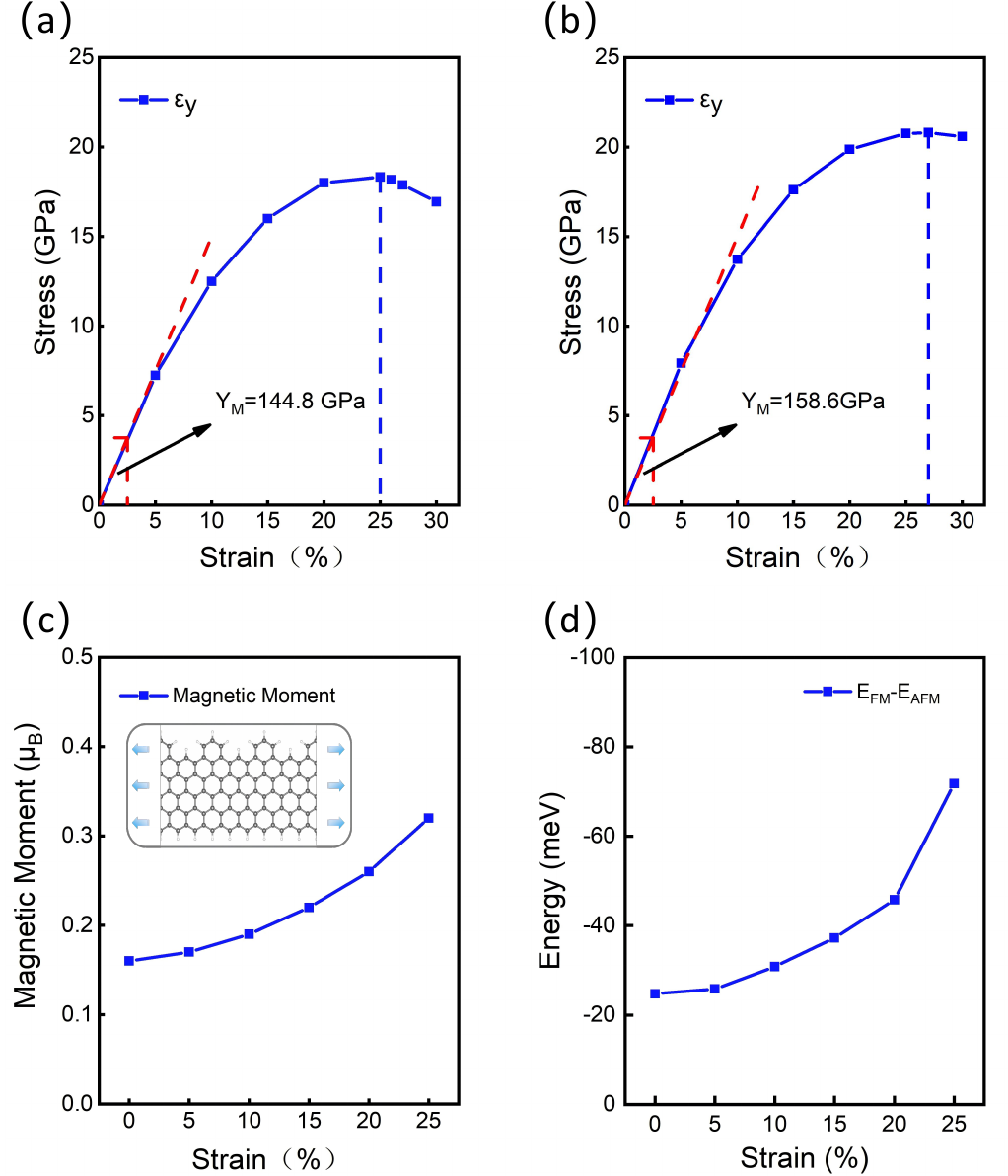} 
\caption{(a) and (b) represent the stress-strain curves for ZGNRs (W=5) and JGNRs (W=5), respectively. (c) shows the strain dependence of the local magnetic moment at the zigzag edge. (d) illustrates the energy difference (FM-AFM) as a function of strain.}
\label{fig:4}
\end{figure}

We then calculated the energy bands, density of states and spin charge densities of the two JGNRs synthesized experimentally for $m = 2$, $W = 4$ and $m = 2$, $W = 5$. It can be seen from the spin charge densities (Fig.\ref{fig:3}(b)) that localized spins do appear at the zigzag edge, while showing negligible spin polarization on one defect flank. It can be seen through the energy band diagram (Fig.\ref{fig:3}(a)) that there is spin polarization in the energy bands of the JGNRs and that the local spins on the side of the zigzag edge are mainly contributed by the spin-up occupied state electrons  near the Fermi energy level. The density of states plot (Fig.\ref{fig:3}(b)) shows that this spin-up occupied state electrons are mainly contributed by the electrons in the $2p_{z}$ orbital of carbon. This is due to the $sp^{2}$ hybridization of the electrons in the plane of graphene, which participates in the formation of $\sigma$ covalent bonds and thus contributes essentially nothing to the system conductivity. Meanwhile, the $pz$ orbitals perpendicular to the plane generate unpaired $\pi$ electrons that primarily govern conduction. Notably, although carbon atoms at jagged edge form $\sigma$ bonds with only two nearest-neighbor carbon atoms in the benzene ring, their three $sp^{2}$ orbitals become saturated through passivation by hydrogen atoms at the edge. Therefore, the $\sigma$ bond band does not contribute to the energy band near the Fermi level, the energy bands near the Fermi level are primarily contributed by the $\pi$ bond bands, which is consistent with the results in the DOS plot. From Fig.\ref{fig:3}(c)-(f), it can also be seen that the conduction band minimum (CBM) states are all contributed by spin-down electrons and the valence band maximum (VBM) states are all contributed by spin-up electrons. This spin-resolved band alignment confirms the system's bipolar magnetic semiconductor character, demonstrating potential for spin-filtering device applications.

Through our calculations and analysis of JGNRs with different widths, we found that the difference between the local magnetic moments of the two structures experimentally synthesized with $m = 2$, $W = 4$, and $m = 2$, $W = 5$ is extremely small, but the structure with $m = 2$, $W = 5$ has a smaller band gap. The reduced bandgap facilitates easier electron excitation and higher carrier mobility. Therefore, we focus on applying uniaxial tensile strain to the $m=2$, $W=5$ JGNRs structure to enhance its intrinsic ferromagnetic properties.

For the structure of JGNRs with $m = 2$ and $W = 5$ we applied uniaxial tensile strains of different magnitudes in the b-direction, the strains are defined as follows: $\varepsilon = \frac{\Delta L}{L_{0}}$. In order to determine the magnitude of the limiting strain in this direction, and estimate the elastic limit of the structure of the JGNRs. We first computed the stress-strain curves of the material Fig.\ref{fig:4}(a), the calculations were made using the method described in references \cite{PhysRevB.64.212103,PhysRevB.66.094110}. This method was originally designed for three-dimensional crystals. In one-dimensional materials, the stress calculated according to the Hellmann-Feynman theorem is adjusted to the force per unit area. It can be clearly seen from the figure that up to 25\% strain, the stress increases gradually with the increase of strain, when it exceeds 25\%, the stress starts to decrease. We therefore determine the maximum sustainable strain to be 25\%. The corresponding tensile strength reaches 18.3 $GPa$. Since within the elastic limit, the stress follows the Hooke's law, $\sigma=Y_{M}\varepsilon$, where $\sigma$ denotes stress, $Y_{M}$ is Young's modulus, and $\varepsilon$ represents strain magnitude \cite{MEMARIAN2015348}. Using this equation, we can calculate the Young's modulus of the material $Y_{M}$ = 144.8 $GPa$. {\color{black}In addition, we also calculated the stress-strain curve of ZGNRs with $W$=5 (Fig.\ref{fig:4}(b)). We determined that the maximum sustainable strain is 27\%. The corresponding tensile strength reaches 20.8 $GPa$, and the Young's modulus $Y_{M}$ = 158.6 $GPa$. The magnitude of Young's modulus of JGNRs ($W$=5) is slightly lower than that of the conventional zigzag graphene nanoribbons with $W = 5$, which is 158.6 $GPa$, this is due to the introduction of topological defects on one of the edges, which leads to a degradation of its mechanical properties.} If we want to increase its mechanical properties experimentally, we can appropriately increase the width of the JGNRs, which can reduce the effect brought by the topological defects.

\begin{figure}[ht]
    \centering
    \includegraphics[width=0.45\textwidth]{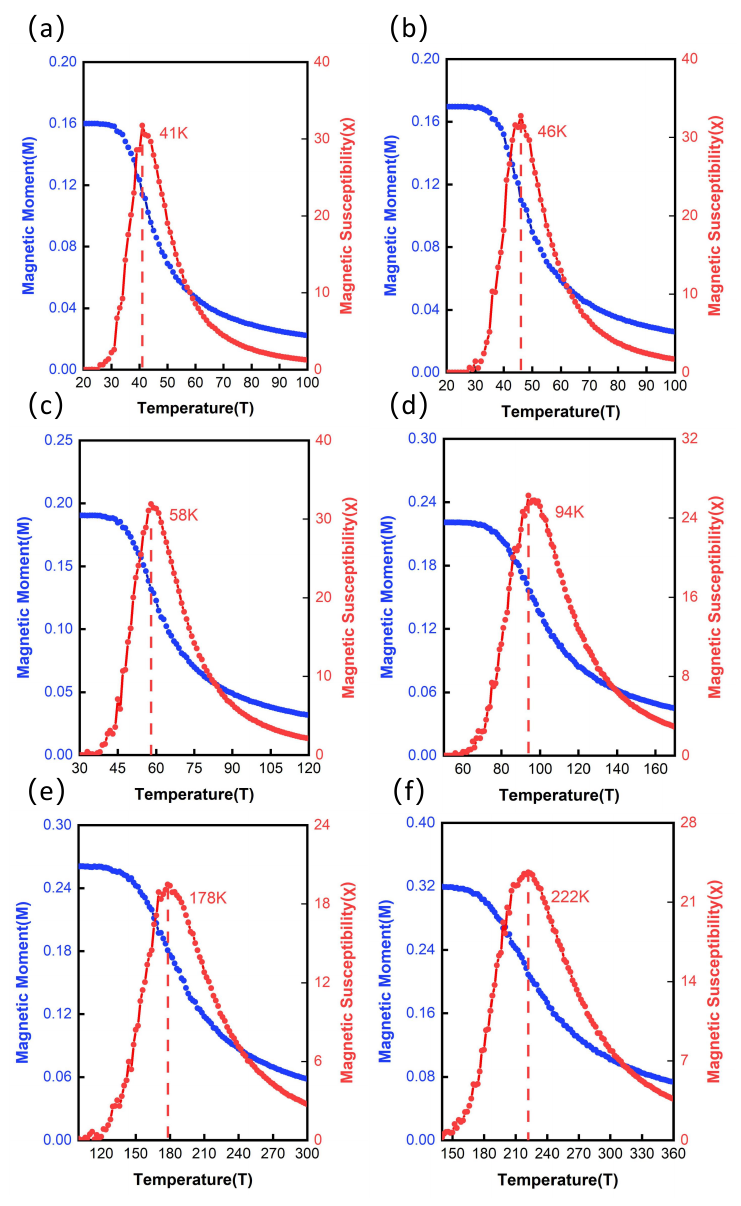} 
    \caption{(a)-(f) Magnetic moments and magnetic susceptibility after applying 0\%, 5\%, 10\%, 15\%, 20\%, and 25\% uniaxial tensile strains to JGNRs with $m = 2$ and $W = 5$, respectively.}
    \label{fig:5}
\end{figure}
\begin{figure}[ht]
    \centering
    \includegraphics[width=0.45\textwidth]{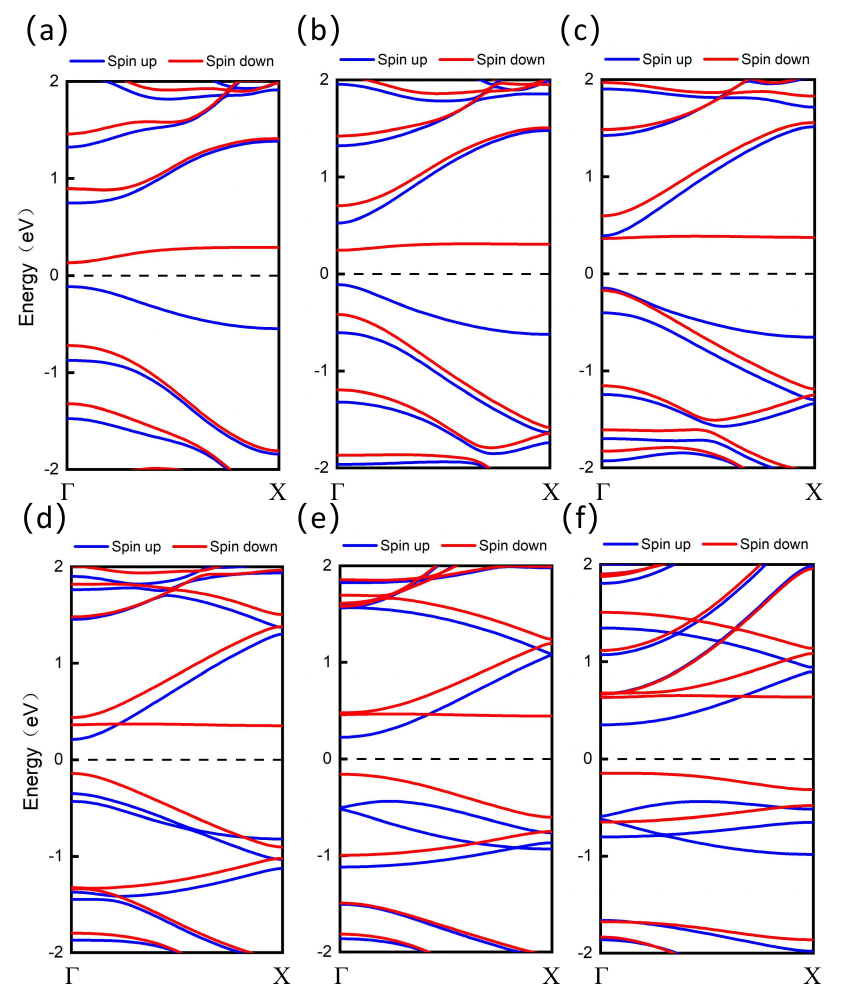} 
    \caption{(a)-(f) Energy bands of spin polarization for JGNRs with $m = 2$ and $W = 5$ after applying 0\%, 5\%, 10\%, 15\%, 20\%, and 25\% uniaxial tensile strain, respectively.}
    \label{fig:6}
\end{figure}

After that, we calculated the local magnetic moments of JGNRs with $m=2$ and $W=5$ at different strains, and we can find from Fig.\ref{fig:4}(c) that with the increase of strain, its local magnetic moment increases from 0.16 $\mu_{B}$ per carbon atoms at no strain to 0.32 $\mu_{B}$ per carbon atoms at 25\% strain. Based on this significant magnetic moment enhancement, we anticipate substantial elevation of the material's Curie temperature. {\color{black}As shown in Fig.\ref{fig:4}(d), the energy difference between the ferromagnetic and antiferromagnetic states gradually increases due to the enhanced magnetic moment and remains negative. This confirms that the material maintains its ferromagnetic ground state under strain, without undergoing a transition to an antiferromagnetic state.}

{\color{black}Subsequently, we employed a Monte Carlo method based on the Ising model to calculate the Curie temperature of the material under different strains. We employ a $1 \times 2 \times 1\ $ supercell structure. Thus the spin Hamiltonian under the Ising model can be written as: 
\begin{equation}
H = - \sum_{i,j} J \, S^z_i S^z_j,
\label{eq:2}
\end{equation}
where the spin $S = 1$ and $J$ is the nearest magnetic exchange coupling parameters. The magnetic
exchange coupling parameters can be calculated through the energy difference between the FM and the AFM configurations, 
\begin{equation}
E_{\text{FM}} = E_0 - 6J|S|^2, \qquad E_{\text{AFM}} = E_0 - 2J|S|^2.
\label{eq:3}
\end{equation}
The energy difference between the ferromagnetic and antiferromagnetic states is given by $\Delta E = E_{AFM }- E_{FM }= 4J|S|^2$. From the table below, we find that $ J>0 $ regardless of the magnitude of the applied strain. Therefore, based on the Ising model, the ground state of this system is ferromagnetic, with an exchange mechanism characterized by direct exchange interaction.

\begin{table}[h!]
    \centering
    \caption{Energy difference, exchange coupling parameters and Curie temperature for different strain levels}
    \label{tab:exchange_coupling_tc}
    \begin{tabular}{cccc}
        \toprule
        Strain & $\Delta E = E_{\text{AFM}} - E_{\text{FM}}$ (\unit{\milli\electronvolt}) & $J$ (\unit{\milli\electronvolt}) & $T_c$ (\unit{\kelvin}) \\
        \midrule
        0\%   & 151.4 & 37.85 & 41 \\
        5\%   & 146.5 & 36.63 & 46 \\
        10\%  & 149.4 & 37.35 & 58 \\
        15\%  & 181.6 & 45.40 & 94 \\
        20\%  & 243.6 & 60.90 & 178 \\
        25\%  & 278.0 & 69.50 & 222 \\
        \bottomrule
    \end{tabular}
\end{table}
} 
When the strain is 0, its intrinsic Curie temperature is 41 $K$. With the increase of strain, the Curie temperature of the material increases significantly, and when the strain reaches the ultimate strain (25\%), its maximum Curie temperature reaches a rare 222 $K$(Fig.\ref{fig:5}(a)-(f)), which is 20 times higher than that of the ZGNRs ($T_{C} < 10 K$), and much higher than the Curie temperature of the partially hydrogenated of graphene nanoribbons ($T_{C}$ = 3.4 $K$)\cite{PhysRevB.83.045414,PhysRevLett.100.047209,DRISSI2015394}. This high Curie temperature ferromagnetism enables JGNRs to be applied as a one-dimensional spin transport device.

In order to probe the microscopic mechanism of the significant increase in the ferromagnetism of this material under strain, we further calculated the energy band structures of JGNRs at different strains, $m=2$, $W=5$, which are displayed in Fig.\ref{fig:6}(a)-(f). The two spin polarized energy bands below the Fermi level of 0.75 $eV$ gradually rise with increasing strain. Meanwhile, as the strain increases, the two spin polarization bands at 0.5 $eV$ above the Fermi level gradually shift downward. When the strain reaches 10\%, the spin-down part of the two spin-polarized bands at 0.75 $eV$ below the original Fermi energy level and the first spin-up band below the original Fermi energy level undergo band degeneracy at the $\gamma$ point. Meanwhile, the spin-up part of the two spin-polarized energy bands at 0.5 $eV$ above the original Fermi energy level and the first spin-down energy band above the original Fermi energy level undergoes band degeneracy at the $\gamma$ point. When strain exceeds 10\%, the two spin-polarized bands below the Fermi energy level will rise further, while the two spin-polarized bands above the Fermi energy level will continue to shift downward. This leads to a significant increase in the density of states near the Fermi level as tensile strain increases. The increase of localized electronic states carrying magnetic moments results in enhanced ferromagnetism within the system.

In addition to this, it is also worth noting that when there is no strain, the VBM is contributed by the spin-up electrons and the CBM is contributed by the spin-down electrons, and when the strain is more than 10\%, the spins of electrons the CBM and the VBM are flipped. The VBM is changed to be contributed by the spin-down electrons and the CBM is changed to be contributed by the spin-up electrons.This characteristic, combined with the gate voltage regulation of the Fermi surface in experiments, can achieve a controllable spin current channel. This enables the material to serve as a novel bipolar magnetic semiconductor material driven by strain. Therefore, this material has become a new candidate for the development and preparation of experimentally tunable spin transport electronic devices\cite{LI2022511,Zhang2020,Mayorov2011}.

{\color{black}Since the PBE functional underestimates the band gap and is sensitive to edge magnetism, we further validated the key results in this paper using the HSE06 functional. For instance, the HSE06 functional reproduces the same qualitative trends as the PBE functional for the band gap, magnetic moment, and the FM-AFM energy difference under strain. This confirms that the trends under strain are robust and not an artifact of the PBE functional's limitations.}

\begin{figure}[ht]
\centering
\includegraphics[width=0.45\textwidth]{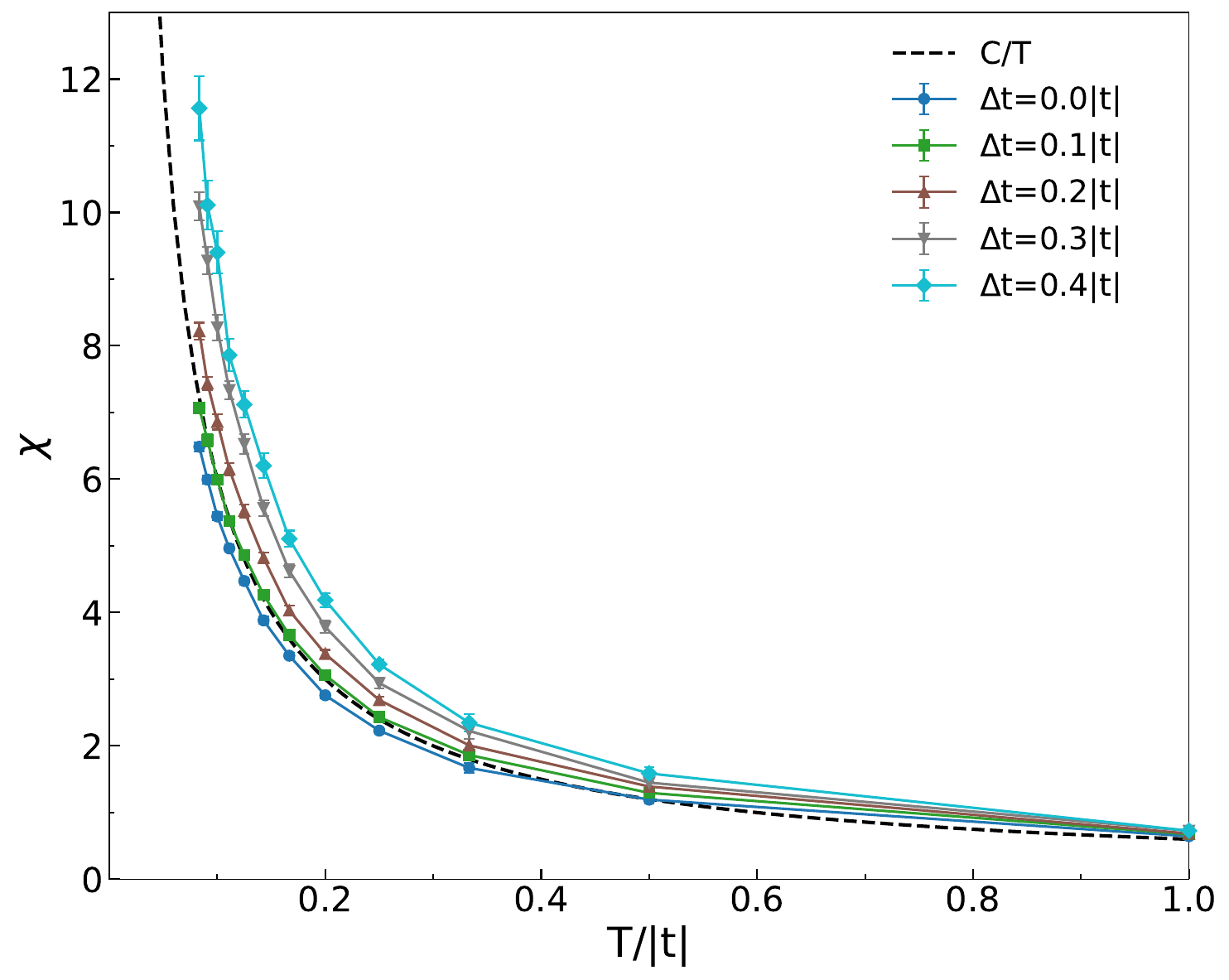} 
\caption{The magnetic susceptibility of zigzag edge $\chi$ at $U = 3.0|t|$ and  $\left\langle n \right\rangle= 1.0$ with different strain.}
\label{fig:7}
\end{figure}

To check our results, we again calculated the magnetic susceptibility ($\chi$) at the zigzag edge of the JGNRs using the DQMC method.
Here
\begin{equation}
\chi = \int_{0}^{\beta} d\tau \sum_{d, d^{\prime} = a, b} \sum_{i, j} \left\langle m_{i_{d}} (\tau) \cdot m_{j_{d^{\prime}}} (0) \right\rangle,
\label{eq:4}
\end{equation}
where $ m_{i_{a}} (\tau) = e^{H \tau} m_{i_{a}} (0) e^{-H \tau} $, with $ m_{i_{a}} = a_{i_{\uparrow}}^{\dagger} a_{i_{\uparrow}} - a_{i_{\downarrow}}^{\dagger} a_{i_{\downarrow}} $
and $ m_{i_{b}} = b_{i_{\uparrow}}^{\dagger} b_{i_{\uparrow}} - b_{i_{\downarrow}}^{\dagger} b_{i_{\downarrow}} $. We measure $ \chi $ in units of $ |t|^{-1} $.

In principle, applying the simplest version of the Hubbard model to correlated electrons in graphene is questionable. However, according to the Peierls-Feynman-Bogoliubov variational principle, the generalized Hubbard model with nonlocal Coulomb interactions can be mapped to an equivalent Hubbard model containing only the effective incumbent interactions $U =1.6|t|$. {\color{black} Following this work, we studied the system's Hamiltonian with $ U=3.0|t| $ and the electron density $ \langle n\rangle $ at half-filling. To fit the Curie-Weiss law, we set the Curie constant $C$ to 0.5.\cite{PhysRevB.94.075106,PhysRevLett.111.036601,PhysRevB.109.075117}}

It is clear from Fig.\ref{fig:7} that $\chi$ is significantly enhanced under strain. {\color{black}This occurs because when the structure undergoes tensile strain, the lattice constant of the system increases, thereby reducing the kinetic energy term $t$ of the electrons. For the $U/t$ ratio, a decrease in $t$ is equivalent to an increase in $U$. Whether $U$ increases or $t$ decreases, electron hopping is suppressed, causing electrons between different lattice points to become more localized. Since the lattice itself possesses ferromagnetic properties, an increase in the number of localized electrons implies enhanced ferromagnetism in the system. Therefore, the ferromagnetism along the zigzag edge of JGNRs increases with increasing strain.}

\section{Conclusion}

In this paper, we theoretically explore the edge ferromagnetism of the recently synthesized Janus graphene nanoribbons and analyze the microscopic mechanism underlying their magnetism, originating from topological defect structures. Building on this foundation, we further investigate the electronic properties and magnetic properties of JGNRs with varying widths and find that their ferromagnetic ground states and spin polarization remain robust, regardless of width. This provides a theoretical basis for synthesizing JGNRs of different widths in future experiments. Notably, applying uniaxial tensile strain further enhances ferromagnetism, raising the Curie temperature to a rare 222 $K$, which surpass previously reported values for ZGNRs and their derivatives. This marks a pivotal step toward realizing room-temperature ferromagnetism in nanoribbons. {\color{black} Indeed, due to the spontaneous symmetry breaking of the $SU(2)$ symmetry at the edge, we expect a gapless Goldstone mode to emerge in the ferromagnetic phase; however, a direct numerical verification of such dynamical excitations goes beyond the scope of the present study and will be left for future investigation.} {\color{black}The combination of micro-nano scale blister technology with atomic force microscopy and micro-Raman spectroscopy represents a promising approach to achieving such large uniaxial tensile strains in JGNRs.\cite{PhysRevLett.119.036101,PhysRevLett.123.116101,Hou2021}} Simultaneously, the emergence of strain-engineered bipolar magnetic semiconductor behavior under uniaxial tension offers a promising pathway for developing next-generation 1D spintronic devices.

\appendix
{\color{black}\section{Magnetic Susceptibility in DFT and DQMC Calculations}
\label{app:chi_calc}

The discrepancy in low-temperature magnetic susceptibility ($\chi$) between density functional theory (DFT) and determinant quantum Monte Carlo (DQMC) calculations arises from their fundamentally different definitions and computational approaches.

In DFT-based calculations, the magnetic susceptibility typically follows the Curie-Weiss law:
\begin{equation}
\chi_{\text{DFT}} = \frac{C}{T - T_c},
\label{eq:chi_dft}
\end{equation}
where $C$ is the Curie constant and $T_c$ is the Curie temperature. According to Eq.~(\ref{eq:chi_dft}), $\chi_{\text{DFT}}$ decreases as $T$ increases above $T_c$. Below $T_c$, $\chi_{\text{DFT}}$ also decreases with decreasing temperature due to the increasing magnitude of $|T - T_c|$. This behavior is physically reasonable for a ferromagnetic system near its ordering temperature.

In contrast, the DQMC-calculated susceptibility is obtained by integrating the imaginary-time correlation function:
\begin{equation}
\chi_{\text{DQMC}} = \int_0^\beta d\tau \sum_{d,d'=a,b} \sum_{i,j} \langle m_{i_d}(\tau) \cdot m_{j_{d'}}(0) \rangle,
\label{eq:chi_dqmc}
\end{equation}
where $\beta = 1/(k_B T)$, $m_{i_d}(\tau) = e^{H\tau} m_{i_d}(0) e^{-H\tau}$ is the imaginary-time evolved magnetization operator at site $i$ on sublattice $d$ ($d = a,b$), and the angular brackets denote thermal averaging. This quantity is often fitted to the form $\chi_{\text{DQMC}} \sim C/T$, which leads to a susceptibility that increases as temperature decreases.

Therefore, the low-temperature $\chi(T)$ curves obtained from the two methods are not expected to be identical due to their distinct theoretical foundations. The DFT-based approach captures the mean-field-like behavior near $T_c$, while DQMC provides a more detailed account of quantum fluctuations and correlations. Crucially, both approaches consistently support the key conclusion that strain enhances edge ferromagnetism in the system.}

\section{ACKNOWLEDGMENTS}
This work was supported by NSFC (12474218) and Beijing Natural Science Foundation (No. 1242022 and 1252022). The numerical simulations in this work were performed at the HSCC of Beijing Normal University.

\nocite{*}
\bibliography{JGNRsReferences}

@article{C8NR00126J,
author ="Liu, Wei and Liu, Jing-yao and Xia, Jing and Lin, Hai-qing and Miao, Mao-sheng",
title  ="Bubble-wrap carbon: an integration of graphene and fullerenes",
journal  ="Nanoscale",
year  ="2018",
volume  ="10",
issue  ="24",
pages  ="11328-11334",
publisher  ="The Royal Society of Chemistry",
doi  ="10.1039/C8NR00126J",
url  ="http://dx.doi.org/10.1039/C8NR00126J"}

@article{C9NR08069D,
author ="Zhao, Lei and Liu, Wei and Yi, WenCai and Hu, Tao and Khodagholian, Dalar and Gu, FengLong and Lin, Haiqing and Zurek, Eva and Zheng, Yonghao and Miao, Maosheng",
title  ="Nano-makisu: highly anisotropic two-dimensional carbon allotropes made by weaving together nanotubes",
journal  ="Nanoscale",
year  ="2020",
volume  ="12",
issue  ="1",
pages  ="347-355",
publisher  ="The Royal Society of Chemistry",
doi  ="10.1039/C9NR08069D",
url  ="http://dx.doi.org/10.1039/C9NR08069D"}

@Article{Novoselov2005,
author={Novoselov, K. S.
and Geim, A. K.
and Morozov, S. V.
and Jiang, D.
and Katsnelson, M. I.
and Grigorieva, I. V.
and Dubonos, S. V.
and Firsov, A. A.},
title={Two-dimensional gas of massless Dirac fermions in graphene},
journal={Nature},
year={2005},
month={Nov},
day={01},
volume={438},
number={7065},
pages={197-200},
issn={1476-4687},
doi={10.1038/nature04233},
url={https://doi.org/10.1038/nature04233}
}

@article{PhysRevLett.100.016602,
  title = {Giant Intrinsic Carrier Mobilities in Graphene and Its Bilayer},
  author = {Morozov, S. V. and Novoselov, K. S. and Katsnelson, M. I. and Schedin, F. and Elias, D. C. and Jaszczak, J. A. and Geim, A. K.},
  journal = {Phys. Rev. Lett.},
  volume = {100},
  issue = {1},
  pages = {016602},
  numpages = {4},
  year = {2008},
  month = {Jan},
  publisher = {American Physical Society},
  doi = {10.1103/PhysRevLett.100.016602},
  url = {https://link.aps.org/doi/10.1103/PhysRevLett.100.016602}
}

@Article{Li2012,
author={Li, Yafei
and Zhou, Zhen
and Shen, Panwen
and Chen, Zhongfang},
title={Electronic and Magnetic Properties of Hybrid Graphene Nanoribbons with Zigzag-Armchair Heterojunctions},
journal={The Journal of Physical Chemistry C},
year={2012},
month={Jan},
day={12},
publisher={American Chemical Society},
volume={116},
number={1},
pages={208-213},
issn={1932-7447},
doi={10.1021/jp207788t},
url={https://doi.org/10.1021/jp207788t}
}

@Article{C5RA06665D,
author ="Chen, Wei and Sun, Yuanhui and Guan, Jia and Wang, Qiang and Huang, Xuri and Yu, Guangtao",
title  ="Molecular charge transfer via π–π interaction: an effective approach to realize the half-metallicity and spin-gapless-semiconductor in zigzag graphene nanoribbon",
journal  ="RSC Adv.",
year  ="2015",
volume  ="5",
issue  ="65",
pages  ="53003-53011",
publisher  ="The Royal Society of Chemistry",
doi  ="10.1039/C5RA06665D",
url  ="http://dx.doi.org/10.1039/C5RA06665D"}

@article{PhysRevB.98.214204,
  title = {Spin-dependent properties in zigzag graphene nanoribbons with phenyl-edge defects},
  author = {Salemi, Leandro and Lherbier, Aur\'elien and Charlier, Jean-Christophe},
  journal = {Phys. Rev. B},
  volume = {98},
  issue = {21},
  pages = {214204},
  numpages = {6},
  year = {2018},
  month = {Dec},
  publisher = {American Physical Society},
  doi = {10.1103/PhysRevB.98.214204},
  url = {https://link.aps.org/doi/10.1103/PhysRevB.98.214204}
}

@Article{C4CP03837A,
author ="Tang, G. P. and Zhang, Z. H. and Deng, X. Q. and Fan, Z. Q. and Zhu, H. L.",
title  ="Tuning spin polarization and spin transport of zigzag graphene nanoribbons by line defects",
journal  ="Phys. Chem. Chem. Phys.",
year  ="2015",
volume  ="17",
issue  ="1",
pages  ="638-643",
publisher  ="The Royal Society of Chemistry",
doi  ="10.1039/C4CP03837A",
url  ="http://dx.doi.org/10.1039/C4CP03837A"}

@Article{Kan2008,
author={Kan, Er-jun
and Li, Zhenyu
and Yang, Jinlong
and Hou, J. G.},
title={Half-Metallicity in Edge-Modified Zigzag Graphene Nanoribbons},
journal={Journal of the American Chemical Society},
year={2008},
month={Apr},
day={01},
publisher={American Chemical Society},
volume={130},
number={13},
pages={4224-4225},
issn={0002-7863},
doi={10.1021/ja710407t},
url={https://doi.org/10.1021/ja710407t}
}

@Article{Mishra2020,
author={Mishra, Shantanu
and Beyer, Doreen
and Berger, Reinhard
and Liu, Junzhi
and Gr{\"o}ning, Oliver
and Urgel, Jos{\'e} I.
and M{\"u}llen, Klaus
and Ruffieux, Pascal
and Feng, Xinliang
and Fasel, Roman},
title={Topological Defect-Induced Magnetism in a Nanographene},
journal={Journal of the American Chemical Society},
year={2020},
month={Jan},
day={22},
publisher={American Chemical Society},
volume={142},
number={3},
pages={1147-1152},
issn={0002-7863},
doi={10.1021/jacs.9b09212},
url={https://doi.org/10.1021/jacs.9b09212}
}

@article{10.1063/1.4747547,
    author = {Bhattacharjee, Joydeep},
    title = {Half-metallicity in graphene nanoribbons with topological defects at edge},
    journal = {The Journal of Chemical Physics},
    volume = {137},
    number = {9},
    pages = {094705},
    year = {2012},
    month = {09},
    issn = {0021-9606},
    doi = {10.1063/1.4747547},
    url = {https://doi.org/10.1063/1.4747547},
}

@Article{Gomes2025,
author={Gomes, Djardiel S.
and Felix, Isaac M.
and Radel, Willian F.
and Dias, Alexandre C.
and Junior, Luiz A. Ribeiro
and Junior, Marcelo L. Pereira},
title={Computational Characterization of the Recently Synthesized Pristine and Porous 12-Atom-Wide Armchair Graphene Nanoribbon},
journal={Nano Letters},
year={2025},
month={May},
day={28},
publisher={American Chemical Society},
volume={25},
number={21},
pages={8596-8603},
issn={1530-6984},
doi={10.1021/acs.nanolett.5c01319},
url={https://doi.org/10.1021/acs.nanolett.5c01319}
}

@article{PhysRevB.110.085141,
  title = {Bipolar magnetic semiconductors emerging in graphene nanoribbons with zigzag edges and internal defects},
  author = {Cheng, Yangkai and Xu, Jing and Xiang, Jiang and Liu, Wei and Miao, Maosheng},
  journal = {Phys. Rev. B},
  volume = {110},
  issue = {8},
  pages = {085141},
  numpages = {11},
  year = {2024},
  month = {Aug},
  publisher = {American Physical Society},
  doi = {10.1103/PhysRevB.110.085141},
  url = {https://link.aps.org/doi/10.1103/PhysRevB.110.085141}
}

@article{
doi:10.1126/science.abq6948,
author = {Song Jiang  and Tomáš Neuman  and Alex Boeglin  and Fabrice Scheurer  and Guillaume Schull },
title = {Topologically localized excitons in single graphene nanoribbons},
journal = {Science},
volume = {379},
number = {6636},
pages = {1049-1054},
year = {2023},
doi = {10.1126/science.abq6948},
URL = {https://www.science.org/doi/abs/10.1126/science.abq6948}}

@Article{Blackwell2021,
author={Blackwell, Raymond E.
and Zhao, Fangzhou
and Brooks, Erin
and Zhu, Junmian
and Piskun, Ilya
and Wang, Shenkai
and Delgado, Aidan
and Lee, Yea-Lee
and Louie, Steven G.
and Fischer, Felix R.},
title={Spin splitting of dopant edge state in magnetic zigzag graphene nanoribbons},
journal={Nature},
year={2021},
month={Dec},
day={01},
volume={600},
number={7890},
pages={647-652},
issn={1476-4687},
doi={10.1038/s41586-021-04201-y},
url={https://doi.org/10.1038/s41586-021-04201-y}
}

@Article{Ruffieux2016,
author={Ruffieux, Pascal
and Wang, Shiyong
and Yang, Bo
and S{\'a}nchez-S{\'a}nchez, Carlos
and Liu, Jia
and Dienel, Thomas
and Talirz, Leopold
and Shinde, Prashant
and Pignedoli, Carlo A.
and Passerone, Daniele
and Dumslaff, Tim
and Feng, Xinliang
and M{\"u}llen, Klaus
and Fasel, Roman},
title={On-surface synthesis of graphene nanoribbons with zigzag edge topology},
journal={Nature},
year={2016},
month={Mar},
day={01},
volume={531},
number={7595},
pages={489-492},
issn={1476-4687},
doi={10.1038/nature17151},
url={https://doi.org/10.1038/nature17151}
}

@Article{Zhang2013,
author={Zhang, Liming
and Yu, Jingwen
and Yang, Mingmei
and Xie, Qin
and Peng, Hailin
and Liu, Zhongfan},
title={Janus graphene from asymmetric two-dimensional chemistry},
journal={Nature Communications},
year={2013},
month={Feb},
day={05},
volume={4},
number={1},
pages={1443},
issn={2041-1723},
doi={10.1038/ncomms2464},
url={https://doi.org/10.1038/ncomms2464}
}

@Article{Lu2017,
author={Lu, Ang-Yu
and Zhu, Hanyu
and Xiao, Jun
and Chuu, Chih-Piao
and Han, Yimo
and Chiu, Ming-Hui
and Cheng, Chia-Chin
and Yang, Chih-Wen
and Wei, Kung-Hwa
and Yang, Yiming
and Wang, Yuan
and Sokaras, Dimosthenis
and Nordlund, Dennis
and Yang, Peidong
and Muller, David A.
and Chou, Mei-Yin
and Zhang, Xiang
and Li, Lain-Jong},
title={Janus monolayers of transition metal dichalcogenides},
journal={Nature Nanotechnology},
year={2017},
month={Aug},
day={01},
volume={12},
number={8},
pages={744-749},
issn={1748-3395},
doi={10.1038/nnano.2017.100},
url={https://doi.org/10.1038/nnano.2017.100}
}

@Article{Zhang2017,
author={Zhang, Jing
and Jia, Shuai
and Kholmanov, Iskandar
and Dong, Liang
and Er, Dequan
and Chen, Weibing
and Guo, Hua
and Jin, Zehua
and Shenoy, Vivek B.
and Shi, Li
and Lou, Jun},
title={Janus Monolayer Transition-Metal Dichalcogenides},
journal={ACS Nano},
year={2017},
month={Aug},
day={22},
publisher={American Chemical Society},
volume={11},
number={8},
pages={8192-8198},
issn={1936-0851},
doi={10.1021/acsnano.7b03186},
url={https://doi.org/10.1021/acsnano.7b03186}
}

@article{10.1063/5.0095203,
    author = {Zhang, Lei and Xia, Yong and Li, Xudong and Li, Luying and Fu, Xiao and Cheng, Jiaji and Pan, Ruikun},
    title = {Janus two-dimensional transition metal dichalcogenides},
    journal = {Journal of Applied Physics},
    volume = {131},
    number = {23},
    pages = {230902},
    year = {2022},
    month = {06},
    issn = {0021-8979},
    doi = {10.1063/5.0095203},
    url = {https://doi.org/10.1063/5.0095203},
}

@Article{Song2025,
author={Song, Shaotang
and Teng, Yu
and Tang, Weichen
and Xu, Zhen
and He, Yuanyuan
and Ruan, Jiawei
and Kojima, Takahiro
and Hu, Wenping
and Giessibl, Franz J.
and Sakaguchi, Hiroshi
and Louie, Steven G.
and Lu, Jiong},
title={Janus graphene nanoribbons with localized states on a single zigzag edge},
journal={Nature},
year={2025},
month={Jan},
day={01},
volume={637},
number={8046},
pages={580-586},
issn={1476-4687},
doi={10.1038/s41586-024-08296-x},
url={https://doi.org/10.1038/s41586-024-08296-x}
}

@article{PhysRevLett.62.1201,
  title = {Two theorems on the Hubbard model},
  author = {Lieb, Elliott H.},
  journal = {Phys. Rev. Lett.},
  volume = {62},
  issue = {10},
  pages = {1201--1204},
  numpages = {0},
  year = {1989},
  month = {Mar},
  publisher = {American Physical Society},
  doi = {10.1103/PhysRevLett.62.1201},
  url = {https://link.aps.org/doi/10.1103/PhysRevLett.62.1201}
}

@article{PhysRevLett.119.076401,
  title = {Topological Phases in Graphene Nanoribbons: Junction States, Spin Centers, and Quantum Spin Chains},
  author = {Cao, Ting and Zhao, Fangzhou and Louie, Steven G.},
  journal = {Phys. Rev. Lett.},
  volume = {119},
  issue = {7},
  pages = {076401},
  numpages = {5},
  year = {2017},
  month = {Aug},
  publisher = {American Physical Society},
  doi = {10.1103/PhysRevLett.119.076401},
  url = {https://link.aps.org/doi/10.1103/PhysRevLett.119.076401}
}

@Article{Jiang2021,
author={Jiang, Jingwei
and Louie, Steven G.},
title={Topology Classification using Chiral Symmetry and Spin Correlations in Graphene Nanoribbons},
journal={Nano Letters},
year={2021},
month={Jan},
day={13},
publisher={American Chemical Society},
volume={21},
number={1},
pages={197-202},
issn={1530-6984},
doi={10.1021/acs.nanolett.0c03503},
url={https://doi.org/10.1021/acs.nanolett.0c03503}
}

@article{PhysRevB.54.11169,
  title = {Efficient iterative schemes for ab initio total-energy calculations using a plane-wave basis set},
  author = {Kresse, G. and Furthm\"uller, J.},
  journal = {Phys. Rev. B},
  volume = {54},
  issue = {16},
  pages = {11169--11186},
  numpages = {0},
  year = {1996},
  month = {Oct},
  publisher = {American Physical Society},
  doi = {10.1103/PhysRevB.54.11169},
  url = {https://link.aps.org/doi/10.1103/PhysRevB.54.11169}
}

@article{KRESSE199615,
title = {Efficiency of ab-initio total energy calculations for metals and semiconductors using a plane-wave basis set},
journal = {Computational Materials Science},
volume = {6},
number = {1},
pages = {15-50},
year = {1996},
issn = {0927-0256},
doi = {https://doi.org/10.1016/0927-0256(96)00008-0},
url = {https://www.sciencedirect.com/science/article/pii/0927025696000080},
author = {G. Kresse and J. Furthmüller}
}

@article{PhysRevB.50.17953,
  title = {Projector augmented-wave method},
  author = {Bl\"ochl, P. E.},
  journal = {Phys. Rev. B},
  volume = {50},
  issue = {24},
  pages = {17953--17979},
  numpages = {0},
  year = {1994},
  month = {Dec},
  publisher = {American Physical Society},
  doi = {10.1103/PhysRevB.50.17953},
  url = {https://link.aps.org/doi/10.1103/PhysRevB.50.17953}
}

@article{PhysRevLett.77.3865,
  title = {Generalized Gradient Approximation Made Simple},
  author = {Perdew, John P. and Burke, Kieron and Ernzerhof, Matthias},
  journal = {Phys. Rev. Lett.},
  volume = {77},
  issue = {18},
  pages = {3865--3868},
  numpages = {0},
  year = {1996},
  month = {Oct},
  publisher = {American Physical Society},
  doi = {10.1103/PhysRevLett.77.3865},
  url = {https://link.aps.org/doi/10.1103/PhysRevLett.77.3865}
}

@article{PhysRevB.13.5188,
  title = {Special points for Brillouin-zone integrations},
  author = {Monkhorst, Hendrik J. and Pack, James D.},
  journal = {Phys. Rev. B},
  volume = {13},
  issue = {12},
  pages = {5188--5192},
  numpages = {0},
  year = {1976},
  month = {Jun},
  publisher = {American Physical Society},
  doi = {10.1103/PhysRevB.13.5188},
  url = {https://link.aps.org/doi/10.1103/PhysRevB.13.5188}
}

@article{ZHANG2021110638,
title = {A universal framework for metropolis Monte Carlo simulation of magnetic Curie temperature},
journal = {Computational Materials Science},
volume = {197},
pages = {110638},
year = {2021},
issn = {0927-0256},
doi = {https://doi.org/10.1016/j.commatsci.2021.110638},
url = {https://www.sciencedirect.com/science/article/pii/S0927025621003657},
author = {Yehui Zhang and Bing Wang and Yilv Guo and Qiang Li and Jinlan Wang}
}

@article{PhysRevD.24.2278,
  title = {Monte Carlo calculations of coupled boson-fermion systems. I},
  author = {Blankenbecler, R. and Scalapino, D. J. and Sugar, R. L.},
  journal = {Phys. Rev. D},
  volume = {24},
  issue = {8},
  pages = {2278--2286},
  numpages = {0},
  year = {1981},
  month = {Oct},
  publisher = {American Physical Society},
  doi = {10.1103/PhysRevD.24.2278},
  url = {https://link.aps.org/doi/10.1103/PhysRevD.24.2278}
}

@article{PhysRevLett.108.066402,
  title = {Metal-to-Insulator Transition and Electron-Hole Puddle Formation in Disordered Graphene Nanoribbons},
  author = {Schubert, Gerald and Fehske, Holger},
  journal = {Phys. Rev. Lett.},
  volume = {108},
  issue = {6},
  pages = {066402},
  numpages = {5},
  year = {2012},
  month = {Feb},
  publisher = {American Physical Society},
  doi = {10.1103/PhysRevLett.108.066402},
  url = {https://link.aps.org/doi/10.1103/PhysRevLett.108.066402}
}

@article{PhysRevB.91.075410,
  title = {Strain-induced edge magnetism at the zigzag edge of a graphene quantum dot},
  author = {Cheng, Shuai and Yu, Jinming and Ma, Tianxing and Peres, N. M. R.},
  journal = {Phys. Rev. B},
  volume = {91},
  issue = {7},
  pages = {075410},
  numpages = {5},
  year = {2015},
  month = {Feb},
  publisher = {American Physical Society},
  doi = {10.1103/PhysRevB.91.075410},
  url = {https://link.aps.org/doi/10.1103/PhysRevB.91.075410}
}

@article{PhysRevB.110.235128,
  title = {Flat bands and superconductivity induced by periodic strain in monolayer graphene},
  author = {Meng, Jingyao and Ma, Runyu and Ma, Tianxing and Lin, Hai-Qing},
  journal = {Phys. Rev. B},
  volume = {110},
  issue = {23},
  pages = {235128},
  numpages = {8},
  year = {2024},
  month = {Dec},
  publisher = {American Physical Society},
  doi = {10.1103/PhysRevB.110.235128},
  url = {https://link.aps.org/doi/10.1103/PhysRevB.110.235128}
}

@Article{2025-0862,
title = {Breathing-Driven Metal-Insulator Transition in Correlated Kagome Systems},
journal = {Chin. Phys. Lett.},
volume = {42},
number = {},
pages = {090712},
year = {2025},
issn = {},
doi = {10.1088/0256-307X/42/9/090712},	
url = {http://cpl.iphy.ac.cn/en/article/doi/10.1088/0256-307X/42/9/090712},
author = {Qingzhuo Duan and Zixuan Jia and Zenghui Fan and Runyu Ma and Jingyao Meng and Bing Huang and Tianxing Ma}
}

@article{PhysRevB.110.085103,
  title = {Disorder-dependent superconducting pairing symmetry in doped graphene},
  author = {Guo, Kaiyi and Zhang, Yue and Liang, Ying and Ma, Tianxing},
  journal = {Phys. Rev. B},
  volume = {110},
  issue = {8},
  pages = {085103},
  numpages = {8},
  year = {2024},
  month = {Aug},
  publisher = {American Physical Society},
  doi = {10.1103/PhysRevB.110.085103},
  url = {https://link.aps.org/doi/10.1103/PhysRevB.110.085103}
}

@article{PhysRevLett.120.116601,
  title = {Localization of Interacting Dirac Fermions},
  author = {Ma, Tianxing and Zhang, Lufeng and Chang, Chia-Chen and Hung, Hsiang-Hsuan and Scalettar, Richard T.},
  journal = {Phys. Rev. Lett.},
  volume = {120},
  issue = {11},
  pages = {116601},
  numpages = {5},
  year = {2018},
  month = {Mar},
  publisher = {American Physical Society},
  doi = {10.1103/PhysRevLett.120.116601},
  url = {https://link.aps.org/doi/10.1103/PhysRevLett.120.116601}
}

@Article{Chen2024,
author={Chen, Chao
and Zhong, Peigeng
and Sui, Xuelei
and Ma, Runyu
and Liang, Ying
and Hu, Shijie
and Ma, Tianxing
and Lin, Hai-Qing
and Huang, Bing},
title={Charge stripe manipulation of superconducting pairing symmetry transition},
journal={Nature Communications},
year={2024},
month={Nov},
day={03},
volume={15},
number={1},
pages={9502},
issn={2041-1723},
doi={10.1038/s41467-024-53841-x},
url={https://doi.org/10.1038/s41467-024-53841-x}
}

@article{PhysRevB.64.212103,
  title = {Ideal strength of diamond, Si, and Ge},
  author = {Roundy, David and Cohen, Marvin L.},
  journal = {Phys. Rev. B},
  volume = {64},
  issue = {21},
  pages = {212103},
  numpages = {3},
  year = {2001},
  month = {Nov},
  publisher = {American Physical Society},
  doi = {10.1103/PhysRevB.64.212103},
  url = {https://link.aps.org/doi/10.1103/PhysRevB.64.212103}
}

@article{PhysRevB.66.094110,
  title = {Ideal strength of bcc molybdenum and niobium},
  author = {Luo, Weidong and Roundy, D. and Cohen, Marvin L. and Morris, J. W.},
  journal = {Phys. Rev. B},
  volume = {66},
  issue = {9},
  pages = {094110},
  numpages = {7},
  year = {2002},
  month = {Sep},
  publisher = {American Physical Society},
  doi = {10.1103/PhysRevB.66.094110},
  url = {https://link.aps.org/doi/10.1103/PhysRevB.66.094110}
}

@article{MEMARIAN2015348,
title = {Graphene Young’s modulus: Molecular mechanics and DFT treatments},
journal = {Superlattices and Microstructures},
volume = {85},
pages = {348-356},
year = {2015},
issn = {0749-6036},
doi = {https://doi.org/10.1016/j.spmi.2015.06.001},
url = {https://www.sciencedirect.com/science/article/pii/S0749603615300239},
author = {F. Memarian and A. Fereidoon and M. {Darvish Ganji}}
}

@article{PhysRevB.83.045414,
  title = {Stability of edge states and edge magnetism in graphene nanoribbons},
  author = {Kunstmann, Jens and \"Ozdo\ifmmode \breve{g}\else \u{g}\fi{}an, Cem and Quandt, Alexander and Fehske, Holger},
  journal = {Phys. Rev. B},
  volume = {83},
  issue = {4},
  pages = {045414},
  numpages = {8},
  year = {2011},
  month = {Jan},
  publisher = {American Physical Society},
  doi = {10.1103/PhysRevB.83.045414},
  url = {https://link.aps.org/doi/10.1103/PhysRevB.83.045414}
}

@article{PhysRevLett.100.047209,
  title = {Magnetic Correlations at Graphene Edges: Basis for Novel Spintronics Devices},
  author = {Yazyev, Oleg V. and Katsnelson, M. I.},
  journal = {Phys. Rev. Lett.},
  volume = {100},
  issue = {4},
  pages = {047209},
  numpages = {4},
  year = {2008},
  month = {Jan},
  publisher = {American Physical Society},
  doi = {10.1103/PhysRevLett.100.047209},
  url = {https://link.aps.org/doi/10.1103/PhysRevLett.100.047209}
}

@article{DRISSI2015394,
title = {Edge effect on magnetic phases of doped zigzag graphone nanoribbons},
journal = {Journal of Magnetism and Magnetic Materials},
volume = {374},
pages = {394-401},
year = {2015},
issn = {0304-8853},
doi = {https://doi.org/10.1016/j.jmmm.2014.08.058},
url = {https://www.sciencedirect.com/science/article/pii/S0304885314007574},
author = {L.B. Drissi and S. Zriouel and E.H. Saidi}
}

@article{LI2022511,
title = {A review of bipolar magnetic semiconductors from theoretical aspects},
journal = {Fundamental Research},
volume = {2},
number = {4},
pages = {511-521},
year = {2022},
issn = {2667-3258},
doi = {https://doi.org/10.1016/j.fmre.2022.04.002},
url = {https://www.sciencedirect.com/science/article/pii/S266732582200173X},
author = {Junyao Li and Xingxing Li and Jinlong Yang}
}

@Article{Zhang2020,
author={Zhang, Mingjia
and Wang, Xiaoxiong
and Sun, Huijuan
and Wang, Naiyin
and He, Jianjiang
and Wang, Ning
and Long, Yunze
and Huang, Changshui
and Li, Yuliang},
title={Induced Ferromagnetic Order of Graphdiyne Semiconductors by Introducing a Heteroatom},
journal={ACS Central Science},
year={2020},
month={Jun},
day={24},
publisher={American Chemical Society},
volume={6},
number={6},
pages={950-958},
issn={2374-7943},
doi={10.1021/acscentsci.0c00348},
url={https://doi.org/10.1021/acscentsci.0c00348}
}

@Article{Mayorov2011,
author={Mayorov, Alexander S.
and Gorbachev, Roman V.
and Morozov, Sergey V.
and Britnell, Liam
and Jalil, Rashid
and Ponomarenko, Leonid A.
and Blake, Peter
and Novoselov, Kostya S.
and Watanabe, Kenji
and Taniguchi, Takashi
and Geim, A. K.},
title={Micrometer-Scale Ballistic Transport in Encapsulated Graphene at Room Temperature},
journal={Nano Letters},
year={2011},
month={Jun},
day={08},
publisher={American Chemical Society},
volume={11},
number={6},
pages={2396-2399},
issn={1530-6984},
doi={10.1021/nl200758b},
url={https://doi.org/10.1021/nl200758b}
}

@article{PhysRevLett.111.036601,
  title = {Optimal Hubbard Models for Materials with Nonlocal Coulomb Interactions: Graphene, Silicene, and Benzene},
  author = {Sch\"uler, M. and R\"osner, M. and Wehling, T. O. and Lichtenstein, A. I. and Katsnelson, M. I.},
  journal = {Phys. Rev. Lett.},
  volume = {111},
  issue = {3},
  pages = {036601},
  numpages = {5},
  year = {2013},
  month = {Jul},
  publisher = {American Physical Society},
  doi = {10.1103/PhysRevLett.111.036601},
  url = {https://link.aps.org/doi/10.1103/PhysRevLett.111.036601}
}

@article{PhysRevB.94.075106,
  title = {Room-temperature magnetism on the zigzag edges of phosphorene nanoribbons},
  author = {Yang, Guang and Xu, Shenglong and Zhang, Wei and Ma, Tianxing and Wu, Congjun},
  journal = {Phys. Rev. B},
  volume = {94},
  issue = {7},
  pages = {075106},
  numpages = {5},
  year = {2016},
  month = {Aug},
  publisher = {American Physical Society},
  doi = {10.1103/PhysRevB.94.075106},
  url = {https://link.aps.org/doi/10.1103/PhysRevB.94.075106}
}

@article{PhysRevB.109.075117,
  title = {Zigzag edge ferromagnetism of triangular-graphene-quantum-dot-like system},
  author = {Han, Runze and Chen, Jiazhou and Zhang, Mengyue and Gao, Jinze and Xiong, Yicheng and Pan, Yue and Ma, Tianxing},
  journal = {Phys. Rev. B},
  volume = {109},
  issue = {7},
  pages = {075117},
  numpages = {8},
  year = {2024},
  month = {Feb},
  publisher = {American Physical Society},
  doi = {10.1103/PhysRevB.109.075117},
  url = {https://link.aps.org/doi/10.1103/PhysRevB.109.075117}
}

@article{PhysRevLett.119.036101,
  title = {Measuring Interlayer Shear Stress in Bilayer Graphene},
  author = {Wang, Guorui and Dai, Zhaohe and Wang, Yanlei and Tan, PingHeng and Liu, Luqi and Xu, Zhiping and Wei, Yueguang and Huang, Rui and Zhang, Zhong},
  journal = {Phys. Rev. Lett.},
  volume = {119},
  issue = {3},
  pages = {036101},
  numpages = {7},
  year = {2017},
  month = {Jul},
  publisher = {American Physical Society},
  doi = {10.1103/PhysRevLett.119.036101},
  url = {https://link.aps.org/doi/10.1103/PhysRevLett.119.036101}
}

@article{PhysRevLett.123.116101,
  title = {Bending of Multilayer van der Waals Materials},
  author = {Wang, Guorui and Dai, Zhaohe and Xiao, Junkai and Feng, ShiZhe and Weng, Chuanxin and Liu, Luqi and Xu, Zhiping and Huang, Rui and Zhang, Zhong},
  journal = {Phys. Rev. Lett.},
  volume = {123},
  issue = {11},
  pages = {116101},
  numpages = {7},
  year = {2019},
  month = {Sep},
  publisher = {American Physical Society},
  doi = {10.1103/PhysRevLett.123.116101},
  url = {https://link.aps.org/doi/10.1103/PhysRevLett.123.116101}
}

@Article{Hou2021,
author={Hou, Yuan
and Dai, Zhaohe
and Zhang, Shuai
and Feng, Shizhe
and Wang, Guorui
and Liu, Luqi
and Xu, Zhiping
and Li, Qunyang
and Zhang, Zhong},
title={Elastocapillary cleaning of twisted bilayer graphene interfaces},
journal={Nature Communications},
year={2021},
month={Aug},
day={20},
volume={12},
number={1},
pages={5069},
issn={2041-1723},
doi={10.1038/s41467-021-25302-2},
url={https://doi.org/10.1038/s41467-021-25302-2}
}

\end{document}